\documentclass[12pt]{mn2e}
\title[The gravitational lens ULAS~J234311.93$-$005034.0]
{ULAS~J234311.93$-$005034.0: A gravitational
lens system selected from UKIDSS and SDSS}
\author[N. Jackson, E.O. Ofek \& M. Oguri]
{Neal Jackson$^{1}$, Eran O. Ofek$^{2}$ and Masamune Oguri$^{3}$\\
$^{1}$Jodrell Bank Centre for Astrophysics, University of Manchester,
Turing Building, Oxford Road, Manchester M13 9PL\\
$^{2}$Division of Physics, Mathematics and Astronomy, California
Institute of Technology, Pasadena, CA 91125\\
$^{3}$Kavli Institute for Particle Astrophysics and Cosmology, Stanford
University, Menlo Park, CA 94025}
\input psfig.tex
\begin{document}
\maketitle
\begin{abstract}
We report the discovery of a new gravitational lens system. This object,
ULAS~J234311.93$-$005034.0, is the first to be selected by using the new 
UKIRT Infrared Deep Sky Survey (UKIDSS), together with the Sloan Digital
Sky Survey (SDSS). The ULAS~J234311.93$-$005034.0 system contains a quasar
at redshift 0.788 which is doubly imaged, with separation 1\farcs4.
The two quasar images have the same redshift and similar, though not 
identical, spectra. The lensing galaxy is detected by subtracting
point-spread functions from $R$-band images taken with the Keck
telescope. 
 The lensing galaxy can
also be detected by subtracting the spectra of the A and B images, since
more of the galaxy light is likely to be present in the latter. No
redshift is determined from the galaxy, although the shape of its
spectrum suggests a redshift of about 0.3. The object's lens status is
secure, due to the identification of two objects with the same
redshift together with a lensing galaxy. Our imaging suggests that the
lens is found in a cluster environment, in which candidate arc-like
structures, that require confirmation, are
visible in the vicinity. Further discoveries of lenses from the 
UKIDSS survey are likely as part of this programme, due to the depth 
of UKIDSS and its generally good seeing conditions.
\end{abstract}

\begin{keywords}
gravitational lensing -- quasars:individual:234311.9$-$005034
\end{keywords}

\section{Introduction}

About 150 gravitational lens systems are now known in which a background
galaxy or quasar is multiply imaged by the action of a foreground
galaxy. Gravitational lens systems are potentially important
because they probe the matter distribution of the lensing galaxy
independently of whether the matter is light emitting (e.g. Cohn et al.
2001; Rusin et al. 2002; Rusin \& Kochanek 2005; Koopmans et al. 2006;
Gavazzi et al. 2007). In cases where the background source is variable,
as in the case of lensed quasars, time delays can also potentially be
used for determination of the Hubble constant according to the method of
Refsdal (1964); for recent reviews see e.g. Courbin (2003), Kochanek \&
Schechter (2004) and Jackson (2007). Lensed quasars can also be used for
studies of quasar structure via microlensing studies (e.g. Agol \&
Krolik 1999; Ofek \& Maoz 2003; Morgan et al. 2007).

For both statistical purposes, and to locate objects important for
particular problems, larger samples are needed. Currently the largest
sample of lenses is the SLACS survey (Bolton et al. 2006), which is
based on the Sloan Digital Sky Survey (SDSS, York et al. 2000) and
relies on selecting SDSS spectra which contain two different redshift
systems at approximately the same spatial position. It is mainly
sensitive to lensed extended galaxies, and has proved to be important in
tackling the problem of mass distributions in lens galaxies (Koopmans et
al. 2006). The other large ($>$20 lens) samples include CLASS, a radio-selected
survey (Myers et al. 2003, Browne et al. 2003) and the SDSS quasar
survey (Inada et al. 2004, Inada et al. 2005, Oguri et al. 2006).
Although individual lenses from these surveys do not always provide good
constraints on the mass distribution in the lensing galaxy, they can be 
used for $H_0$ determination and also for statistical studies (e.g. 
Chae et al. 2003; Ofek, Rix \& Maoz 2003; Mitchell et al. 2005; Oguri
2007; Oguri et al. 2008).

As already demonstrated, the 
Sloan Digital Sky Survey (SDSS, York et al. 2000) is potentially a
very important resource, due to its wide-area sky coverage and
consequent large number of quasars. The latest catalogue from
the SDSS (Schneider et al. 2005) contains 77429 quasars. With typical
lensing rates (e.g. Turner, Ostriker \& Gott 1984) 
this might be expected to yield more than 
100 gravitational lenses. Unfortunately, many of these lenses will
have relatively small separations, as high-resolution searches with
radio interferometers reveal a median separation of between 1\arcsec and
1\farcs5, and substantial numbers with separations smaller than
1\arcsec. The combination of seeing effects (SDSS imaging has a median
PSF width of 1\farcs4 FWHM) and weak secondary images mean that
moderate- to small-separation lenses will be difficult to detect
directly. Nevertheless, 24 SDSS quasar lenses are known (Oguri et al. 
2006, Inada et al. 2007) from searches at larger separation. At smaller
separation attempts have been made to overcome the difficulties,
including the use of colour selection to identify candidates which
contain light from both quasar and lens galaxy (Ofek et al. 2007) and
the use of radio catalogues together with optical information from SDSS
(Jackson \& Browne 2007). Here we report an alternative approach, using
new large-area public infrared surveys.

The UK Infrared Telescope (UKIRT) is currently carrying out a number of large
surveys in multiple (YJHK) infra-red bands, some in the areas surveyed by the
SDSS,  with the collective name of the UKIRT
Infrared Deep Sky Survey (UKIDSS). 
The UKIDSS project is defined in Lawrence et al (2007). UKIDSS uses the
UKIRT Wide Field Camera (WFCAM; Casali et al, 2007) and a photometric
system described in Hewett et al (2006). The pipeline processing and
science archive are described in Irwin et al (2008, in preparation) and Hambly
et al (2008).

Of particular interest here is the Large Area
Survey (LAS), which will eventually cover 4028 square degrees to a depth
of $K$=18.2 in seeing of $<$1\farcs2 and with median seeing below
1\arcsec. This 25--50\% improvement over SDSS seeing is vital for 
the discovery of a gravitational lens whose separation is
just smaller than the SDSS point spread
function. Moreover, the colour information using the combination of
optical and infrared bands may also be important in prioritising
candidates and deciding on their lensing nature. This paper presents the
first results of an attempt to use SDSS and UKIDSS together for lens
discovery, and further work will extend this to the entire SDSS quasar
sample. Here we report on the selection of a joint UKIDSS-SDSS sample,
resulting in discovery of a probable gravitational
lens, ULAS~J234311.93$-$005034.0. In Section 2 we describe the candidate
selection; in section 3 we describe follow-up observations with Keck
imaging and spectroscopy of  ULAS~J234311.93$-$005034.0, and briefly
discuss the results in section 4.

\section{Selection}

We used the sample of 77429 quasars from the SDSS/Data Release 5 (DR5, 
Schneider et al. 2007); this data release has a footprint of
approximately 8000 square degrees. Within 5$^{\prime\prime}$ of each object we
searched for UKIDSS sources using the $H$-band catalogue from the UKIDSS Large
Area Survey Data Release 2 (Warren et al. 2007) which has a footprint of
685 square degrees. 
6707 objects are found 
which have cross-matches in the UKIDSS survey, which represent the vast 
majority of SDSS quasars within the UKIDSS DR2 area. The UKIDSS DR2 area
covers approximately $1\deg$ either side of the equator from right
ascensions of approximately 0$^{\rm h}$-3$^{\rm h}$30$^{\rm m}$ and 22$^{\rm
h}$30$^{\rm m}$-24$^{\rm h}$, with smaller, mostly equatorial fields between
08$^{\rm h}$ and 16$^{\rm h}$.

UKIDSS $H$-band images of all of these cross-matched objects were then 
individually inspected by eye. Objects were selected for further investigation
if they appeared to have extensions or secondary structure within approximately
2\arcsec, as this is the range within which selection using SDSS alone
is likely to become incomplete. In cases of doubt, all available UKIDSS
wavelength bands were scrutinised to check that secondary structure was
visible in more than one colour. No distinction was made on relative
colour of primary and secondary, as it is possible that the light of one
of the images is dominated by light from the lensing galaxy, which is
expected to be closer to the weaker image.

From the initial sample of 6707 quasars, 56 candidates were identified as
possible gravitational lenses. Of these, the one known lens
(J122608.0$-$000602, Inada et al. 2008), is recovered as a
candidate\footnote{The footprint of the subsequent release, DR3, of
UKIDSS contains three known lenses, all of which were recovered as candidates.}.
Two other objects were in the list of candidates
selected on the basis of SDSS colours by Oguri et al. (2007), one of
which (J001342.4$-$002413) has already been ruled out by followup
observations.

\section{ULAS J234311.9$-$005034}

One of the promising candidates identified in the selection
by eye was ULAS~J234311.93$-$005034.0, a double source with approximate 
separation of 1\farcs4. This object has a K-band magnitude of 16.5 and an
optical ``best-PSF'' $r$-band magnitude of 20.3, from the UKIDSS and SDSS surveys
respectively. Selected UKIDSS and SDSS ($g$-band) images are shown together in
Figure 1. The object lies in the FIRST 20-cm radio survey area (Becker
et al. 1995), but no radio counterpart is detected at a level of 
$\geq$1~mJy (5$\sigma$).

\begin{figure}
\begin{tabular}{c}
\psfig{figure=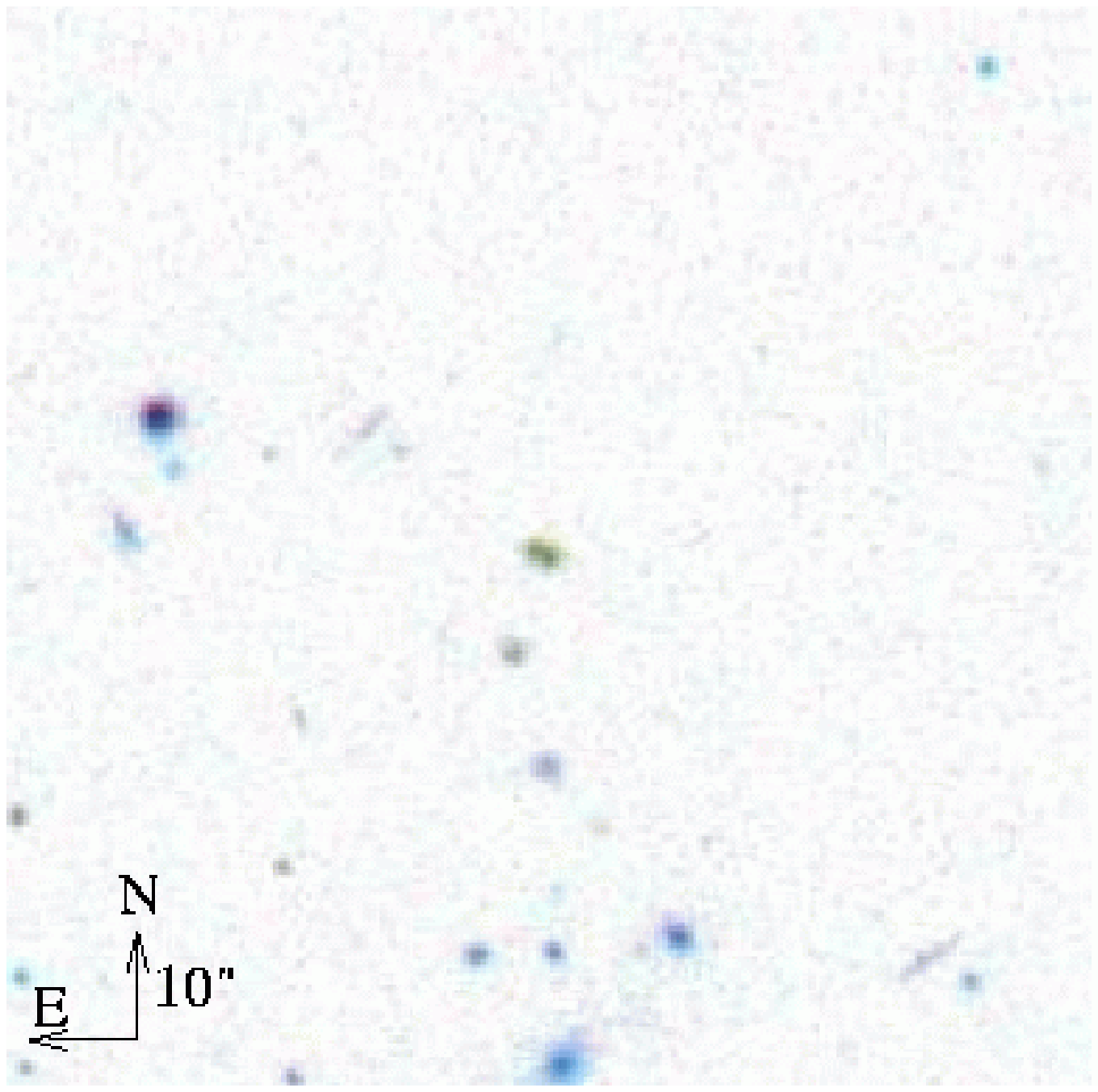,width=8cm}\\
\psfig{figure=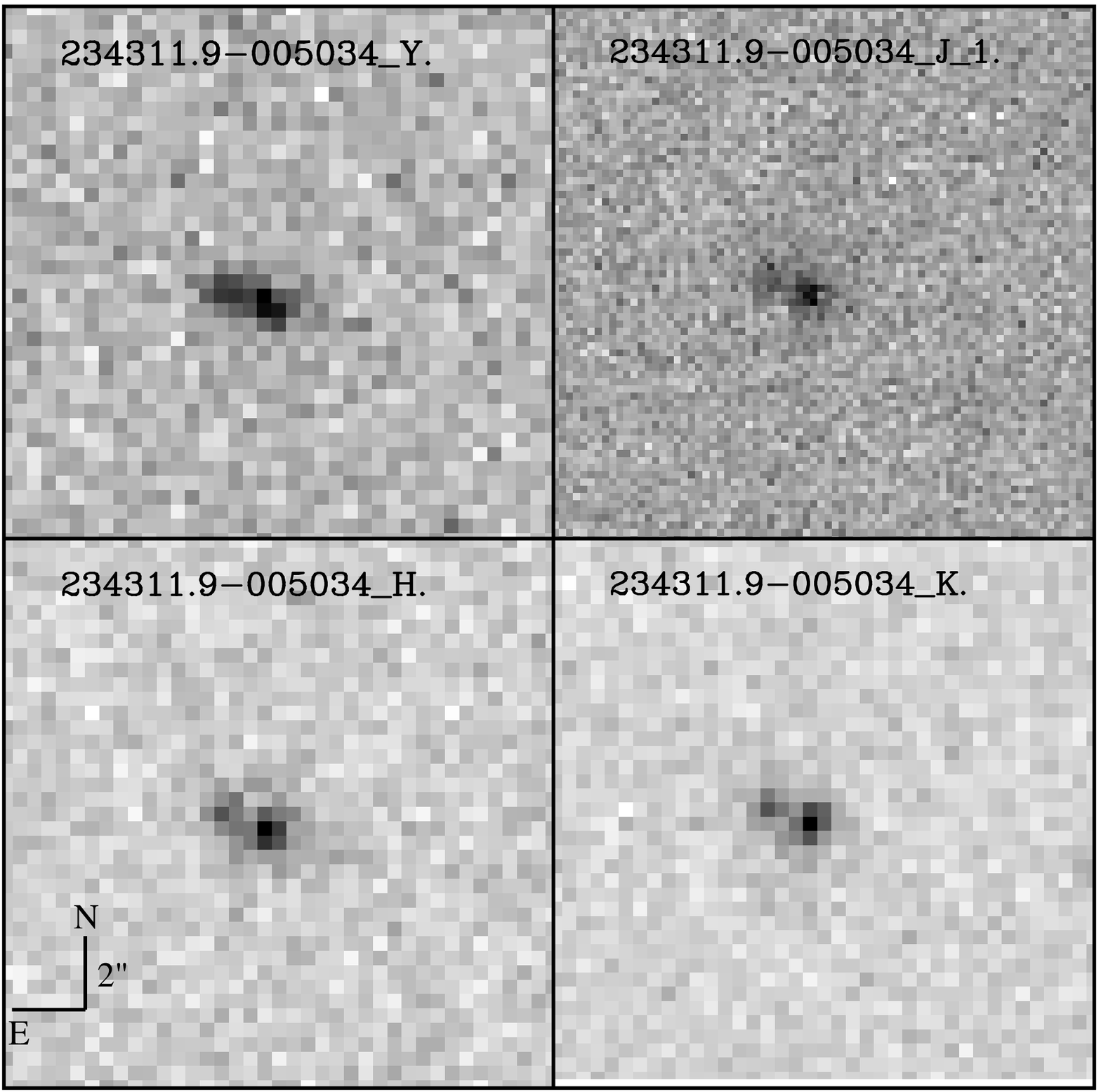,width=8cm}\\
\end{tabular}
\caption{Top: SDSS $g$ image of the field of ULAS~J234311.93$-$005034.0. The
image is 101\arcsec on a side and shows the object as slightly extended
in a direction North of East. Bottom: UKIDSS images in four colours ($Y$,
$J_1$, $H$ and $K$). Each image is 15\arcsec on a side.}
\end{figure}

\subsection{Keck imaging}

Follow-up imaging and spectroscopy was performed using the Keck-I telescope
and the Low-Resolution Imaging Spectrograph -- Atmospheric
Dispersion Compensator (LRIS--ADC; Oke et al. 1995) 
on the night of 2008 January 04. Imaging exposures totalling 1140~s were
obtained in the $R$ filter, and 1500~s in the $g$ filter. The
seeing at the time, measured from nearby stars, was about 1\arcsec. 
Fig. 2 shows the Keck images in both filters. In both figures, the object is
clearly visible as a double source, confirming the impression from the
UKIDSS survey images. 

\begin{figure}
\psfig{figure=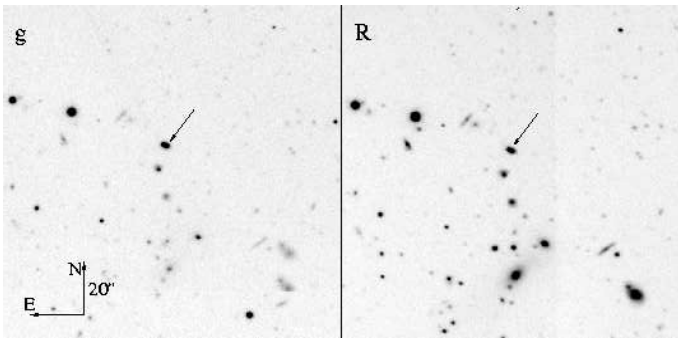,width=9cm,angle=-90}
\caption{Keck images in the $g$-band (left) and $R$-band (right).}
\end{figure}

The $R$-band image was fitted using the {\sc galfit} software of Peng et
al. (2002) using two point spread functions, derived from a nearby
isolated and unsaturated stellar image, a Sersic profile, and a flat
background to represent the sky. The
positions and fluxes of the two PSF images were allowed to vary, as were
the position, flux, position angle, effective radius and Sersic index of
the galaxy. The resulting fits are shown in Fig. 3, and give an image
separation of 1\farcs4 between two point sources differing by about 20\%
in flux. It can be seen, by
comparing the raw image and the fitted points, that additional flux is
required in excess of the point sources in order to fit the sources
properly. According to the fit, this can be provided by a galaxy
slightly brighter than either image, whose major axis lies close to the
line joining the two images. Table 1 shows the fitted parameters of the
galaxy and lensed images.

\begin{figure}
\begin{tabular}{cc}
\psfig{figure=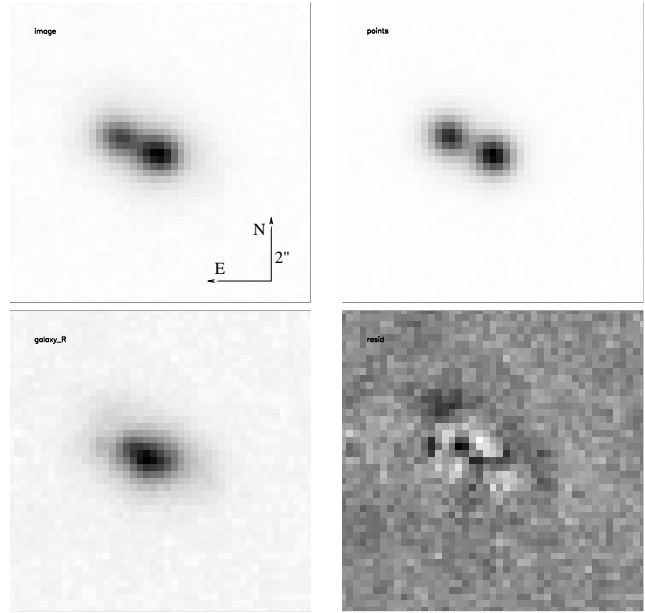,width=4cm}&
\psfig{figure=R_points.fits.ps,width=4cm,angle=-90}\\
\psfig{figure=R_fitgal.fits.ps,width=4cm,angle=-90}&
\psfig{figure=R_resid.fits.ps,width=4cm,angle=-90}\\
\end{tabular}
\caption{Zoom-in on the $R$-band image of the lensed quasars
ULAS~J234311.93$-$005034.0. (Top left) the image of the 
ULAS~J234311.93$-$005034.0 region; (top right) the fitted PSFs 
for images A and B; (bottom left) the fitted galaxy; and (bottom 
right) the residual (image minus the fitted PSF and galaxy components).}
\end{figure}

A similar attempt was made to fit the $g$-band image, but the fit fails
to represent both the lensed images and the galaxy. This is likely to be
because the galaxy is relatively weaker, and the fit collapses to a
state in which both the galaxy and one of the images are at the same
position. However, if we constrain the position of the centre of the
galaxy in the $g$-band to be the same as that for the $R$-band, we
obtain a good fit in which the fitted position angle of the galaxy
agrees well with that in the $R$-band image. 
Accordingly, Table 1 presents this fit to the
$g$-band image. The discrepancy between the two
colour bands is likely to be due to the large errors resulting from the
attempt to fit components separated
from each other by distances slightly larger than the seeing FWHM.


\begin{table}
\begin{tabular}{cccc}
&&$R$-band&$g$-band\\
Image A &Magnitude &20.60  & 20.35\\
        &Position /\arcsec  &(0.0,0.0)&(0.0,0.0)\\
Image B &Magnitude &20.78  & 20.91\\
        &Position /\arcsec  &(1.39,0.59)&(1.21,0.51)\\
Galaxy  &Magnitude &19.94  & 21.83\\
        &Position /\arcsec &(0.40,0.20)  & (0.40,0.20)$^*$\\
        &PA &72  & 68\\
        &Axis ratio $b/a$ &0.43  & 0.70\\
        &$r_e$/\arcsec &0\farcs94  &1\farcs54 \\
        &Sersic index &1.56  & 0.48\\
\end{tabular}
\caption{Fit parameters for the Keck imaging of ULAS~J234311.93$-$005034.0. All parameters
are given with respect to the brighter image A, including position
offsets in arcseconds (positive values indicate increasing RA and Dec).
Position errors from the fit are nominally 2--3~mas, but are likely to
be larger, particularly for the galaxy position. The galaxy position in
the $g$-band fit, marked with an asterisk, 
is fixed with respect to image A. Magnitudes are determined in GALFIT by
the fitting of a PSF to the images, and a Sersic law for the galaxy. 
The photometric scale is
calibrated by use of Sloan $g,r,i$ values for this object and the
transformation equations of Jester et al. (2005).}
\end{table}

Although the quasar is included in the SDSS DR5 quasar catalogue
(Schneider et al. 2007), its SDSS magnitude of $i=20.1$, after extinction
correction, is fainter than the magnitude limit of the SDSS quasar lens
survey (SQLS, Inada et al. 2003, 2005, 2008; 
Oguri et al. 2006) and is therefore not part of the statistical
SQLS sample.

\subsection{Keck spectroscopy}

Spectra were obtained, also using the Keck-I telescope, on the night of 2008
January 11. For the spectroscopy the 400/3400 grism on the blue arm of
LRIS (central wavelength 440~nm), together with the 400/8500 grating on 
the red arm (central wavelength 770~nm) and 
a dichroic cutting at about 560~nm, were used. The total exposure time
was 1400~s, and a 0\farcs 7 slit was used throughout. 
Data reduction followed standard
procedures including bias subtraction with 10 coadded bias frames and
flatfielding from flat-field frames made using a halogen lamp.
The extracted, sky-subtracted spectra are shown in Fig.~4. The 
two QSO images clearly have the same redshift,
which is determined to be 0.788$\pm$0.001 using the narrow [O{\sc iii}]
line at 500.7~nm. The spectra are, however, not identical. If we use the
fit to the $R$-band image described above to give the ratio of the
fluxes of the quasar images, we can recover a spectrum of the galaxy
alone; this is also shown in Fig.~4. This is obviously redder than the
quasar images, but there is no obvious emission or absorption line. The
general shape of the spectrum suggests an upturn due to a 4000\AA\ break
at around 5500\AA, implying a redshift of about 0.3; in this case,
deeper exposures would be needed to detect the Ca H/K absorption
lines.

\begin{figure}
\psfig{figure=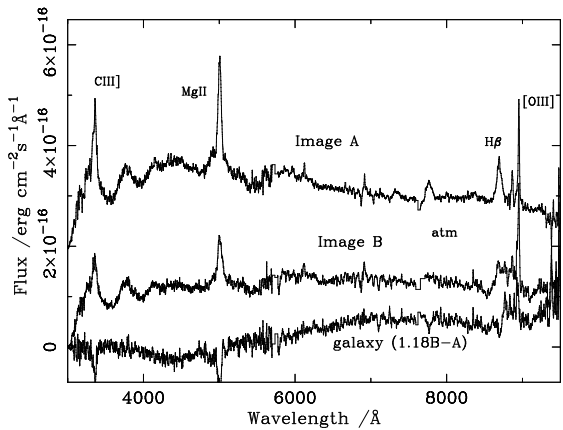,width=8cm,angle=-90}
\caption{Spectra of images A and B of the ULAS~234311.93$-$005034.0 system.
Note that the spectrum of image A has been displaced upwards by 
2$\times$10$^{-16}$erg\,cm$^{-2}$s$^{-1}$\AA$^{-1}$ for clarity. Cosmic
rays and telluric features have been removed.}
\end{figure}

\section{Discussion and conclusions}

The detection of two images with the same redshift, together with clear
evidence for a lensing galaxy, convinces us that
ULAS~J234311.93$-$005034.0 is a lens system. Further circumstantial
evidence can be obtained by considering the expected brightness of the
lens galaxy from the image separation and the Faber-Jackson relation 
(Faber \& Jackson 1976). If we use the lens separation together with 
the Faber-Jackson relation as calibrated for a sample of gravitational 
lenses by Rusin et al. (2003), and assume a lens redshift of 0.32, we 
obtain an expected $R$-band magnitude of 19.7, very close to the actual 
value.

Nevertheless, it is obvious that the spectra (Fig. 4) are not identical,
despite the very similar redshift. The continuum shape is different, but
as we have seen it is plausible that this is due to different levels of
contamination from the lensing galaxy. A complicating factor, however, 
is that there may be differential reddening between
images A and B. This can be deduced from ratios of the broad emission 
line fluxes
between image A and B, which decrease systematically with wavelength
(Table 2). If this reddening takes place in the lensing galaxy, and if
its redshift is about 0.3, this implies an associated $E(B-V)$ of about
0.7 for a Galactic extinction law.

\begin{table}
\begin{tabular}{lccc}
Line/rest wavelength & Image A & Image B & Ratio (A/B)\\ \hline
C{\sc iii}] 190.9~nm (narrow) & 7.2 & 3.3 & 2.19 \\
C{\sc iii}] 190.9~nm (broad)  & 26.4 & 20.0 & 1.32 \\
Mg{\sc ii} 279.8~nm (narrow) & 9.7 & 5.2 & 1.86 \\
Mg{\sc ii} 279.8~nm (broad) & 9.5 & 6.3 & 1.51 \\
H$\beta$ 486.1~nm & 6.4 & 9.0 & 0.72 \\
\mbox{}[O{\sc iii}] 500.7~nm & 4.2 & 6.2 & 0.68\\ \hline
\end{tabular}
\caption{Line fluxes from images A and B separately, and the flux ratio 
A/B, measured by
subtracting a polynomial fit to the continuum by eye and fitting
Gaussians. All fluxes are in units of 10$^{-15}$erg\,cm$^{-2}$s$^{-1}$
and are likely to be subject to errors of between 5 and 10\%.
The broad component of Mg{\sc ii} in particular probably contains large
amounts of contamination from Fe{\sc ii}.}
\end{table}

Assuming that the reddening occurs in the lensing galaxy, it is then
puzzling why the fit to the Keck images places the lensing
galaxy closer to A than to B. There are precedents for this (e.g.
CLASS~B0218+357; Jackson, Xanthopoulos \& Browne 2000, HE1104$-$185;
Wisotzki et al. 1993) but an
alternative explanation is that the fitted centroid of the galaxy is in
error. HST imaging would be useful to resolve
this matter fully. Reddening of the quasar continuum is difficult to
assess, due to contamination of both A and B spectra by light from the
lensing galaxy. An additional complication is the possibility of
microlensing in the lensing galaxy, which may affect both the quasar
continua and, if the broad line region is small enough, the broad line
fluxes. It is becoming increasingly clear (e.g. Kochanek et al. 2007,
Morgan et al. 2006, Keeton et al. 2006, Gaynullina et al. 2005) that
microlensing is important in understanding relative fluxes in images of
lensed quasars.

It is possible to make a lens model to describe the system, although it
suffers from a lack of constraints. The available constraints are the
image positions and fluxes (assuming we ignore effects of reddening or
microlensing),
and the galaxy position is known. We assume that positions are known to
10~mas and fluxes to 10\%, and use a singular isothermal ellipsoid to
represent the mass profile of the galaxy. Fixing the ellipticity at 0.5 and the
position angle at the same as that of the light distribution, but
varying the lens Einstein radius, gives $\chi^2/d.o.f.$=595/2. If in
addition we allow the ellipticity to vary, we obtain $\chi^2/d.o.f.$=48/1
at the cost of an unlikely solution of an almost circular galaxy. If we
once more fix the ellipticity at 0.5, but allow two extra free
parameters to represent external shear, we remove all the degrees 
of freedom and therefore obtain an exact fit (Fig. 5). This fit
has a very large shear magnitude (0.34) at a position angle of
$-18^{\circ}$, and predicts a time delay of 18 days for
$H_0$=70~kms$^{-1}$Mpc$^{-1}$, with the A image varying first. 
Interestingly, the predicted shear direction is within a few degrees of 
the direction of the nearest object, a galaxy about 10\arcsec\ away and 
slightly west of south. 

\begin{figure}
\psfig{figure=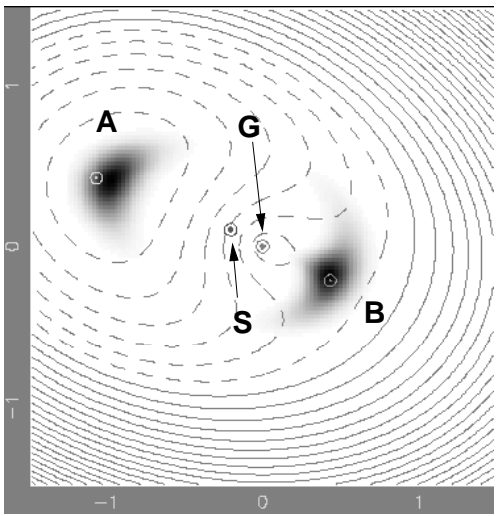,width=6.5cm}
\caption{Straw-man lens model for the ULAS~J234311.93$-$005034.0 system.
The scale on $x$ and $y$ axes is in arcseconds. Images A and B are
marked, together with the centre of the galaxy (G) and the model source
position (S). The greyscale shows the
predicted images and the contours are time-delay contours. The measured
positions of the images (at either end) and lensing galaxy (second from
right) are shown, as is the inferred position of the source. The model
includes external shear of magnitude 0.34 and direction $-$18$^{\circ}$,
and has a fixed ellipticity of 0.5 in PA~72$^{\circ}$. The predicted
Einstein radius of the lens galaxy is 0\farcs58.}
\end{figure}

That there may be an effect of the environment on the lens model is
consistent with the appearance of the Keck images. The
ULAS~J234311.93$-$005034.0 system appears to lie at the northern end of a
cluster of galaxies, the brightest one of which lies about 49\arcsec\
South. The redshift of this bright galaxy appears in the SDSS database
as 0.316, so it is at least plausible that the ULAS~J234311.93$-$005034.0
lens galaxy redshift lies at around this value. 
There are a a number of arc-like structures in this cluster,
suggestive of weak lensing of background objects (see Fig.~2). The
velocity dispersion of the cluster implied by the presence of a shear 
of 0.34 at the position of the gravitational lens is large
($\sim$1400~km$\,$s$^{-1}$ assuming an isothermal profile for the cluster)
but detailed modelling with additional constraints is needed in order to
make a serious estimate of the cluster mass.

The discovery of ULAS~J234311.93$-$005034.0 suggests that the use of
large-area surveys with good spatial resolution is a good way to
discover lenses in existing surveys of quasars such as the SDSS quasar
sample. The increasing coverage of the UKIDSS survey means that
ultimately between 60000-70000 high-resolution images of quasars should
be available. For typical angular separation distributions of existing
lens samples, this should allow the existing sample of 24 SDSS quasar
lenses to be expanded by at least a factor 2, and possibly more.

\normalsize

\section*{References}

\parindent 0mm

Agol E., Krolik J., 1999, ApJ 524, 49

Becker R.H., White R.L., Helfand D.J., 1995, ApJ 450, 559

Bolton A.S., Burles S., Koopmans L.V.E., Treu T., Moustakas L.A. 2006,  ApJ, 638, 703. 

Browne I.W.A. et al. 2003,  MNRAS, 341, 13. 

Casali M., et al., 2007, A\&A 467, 777.

Chae K.H. 2003,  MNRAS, 346, 746. 

Cohn J.D., Kochanek C.S., McLeod B.A., Keeton C.R. 2001,  ApJ, 554, 1216. 

Courbin F., 2003, astro-ph/0304497.

Faber S.M., Jackson R.E., 1976, ApJ, 204, 668

Gavazzi R., Treu T., Rhodes J.D., Koopmans L.V.E., Bolton A.S., Burles S., Massey R.J., Moustakas L.A. 2007,  ApJ, 667, 176. 

Gaynullina E.R., et al., 2005, A\&A 440, 53

Hambly N., et al., 2008, MNRAS 384, 637

Hewett P.C., Warren S.J., Leggett S.K., Hodgkin S.T., 2006, MNRAS,  367, 454

Inada N., et al., 2003, Nature 426, 810

Inada N., et al. 2005,  AJ, 130, 1967. 

Inada N., Oguri M., Becker R.H., White R.L., Kayo I., Kochanek C.S., Hall P.B., Schneider D.P., York D.G., Richards G.T. 2007,  AJ, 133, 206. 

Inada N., et al. 2008, AJ 135, 496

Jackson N., Browne I.W.A. 2007,  MNRAS, 374, 168. 

Jackson N., Xanthopoulos E., Browne I.W.A. 2000,  MNRAS, 311, 389. 

Jackson N. 2007,  LRR, 10, 4. 

Jester S., et al., 2005, AJ, 130, 873.

Keeton C.R., Burles S., Schechter P.L., Wambsganss J., 2006, ApJ 639, 1
 
Kochanek C.S., Schechter P.L. 2004, Measuring and Modelling the
Universe, Carnegie Obs. Centennial Symposium, ed. W. Freedman, CUP, p.117. 

Kochanek C.S., Dai X., Morgan C., Morgan N., Poindexter S., in
Statistical Challenges in Modern Astronomy IV, 2007, eds Babu G.J. et al.,
San Francisco:ASP.

Koopmans L.V.E., Treu T., Bolton A.S., Burles S., Moustakas L.A. 2006,  ApJ, 649, 599. 

Lawrence A., et al. 2007,  MNRAS, 379, 1599. 

Mitchell J.L., Keeton C.R., Frieman J.A., Sheth R.K. 2005,  ApJ, 622, 81. 

Morgan C.W., et al., 2007, ApJ submitted, astro-ph/0710.2552

Morgan C.W., Kochanek C.S., Morgan N.D., Falco E.E., 2006, ApJ 647, 874

Myers S.T., et al. 2003,  MNRAS, 341, 1. 

Oke J.B., et al., 1995, PASP 107, 375

Ofek E.O., Maoz D., 2003, ApJ 594, 101

Ofek E.O., Rix H.-W., Maoz D., 2003, MNRAS 343, 639

Ofek E.O., Oguri M., Jackson N., Inada N., Kayo I. 2007, MNRAS, 382, 412.

Oguri M., et al. 2006,  AJ, 132, 999. 

Oguri M. 2007,  ApJ, 660, 1. 

Oguri M. et al. 2008, AJ, 135, 512

Peng C.Y. Ho L.C., Impey C.D., Rix H. 2002, AJ 124, 266.

Refsdal S. 1964,  MNRAS, 128, 307. 

Rusin D., Kochanek C.S. 2005,  ApJ, 623, 666. 

Rusin D., Norbury M., Biggs A.D., Marlow D.R., Jackson N.J., Browne I.W.A., Wilkinson P.N., Myers S.T. 2002,  MNRAS, 330, 205. 

Rusin D., et al., 2003, ApJ, 587, 143

Schneider D.P., et al. 2007,  AJ, 134, 102. 

Turner E.L., Ostriker J.P., Gott J.R., 1984, ApJ, 284, 1.

Warren S.J., et al. 2007,  MNRAS, 375, 213. 

Wisotzki L., Koehler T., Kayser R., Reimers D., 1993, A\&A, 278, L15

York D.G., et al. 2000,  AJ, 120, 1579. 

\section*{Acknowledgements}

We would like to thank the Kavli Institute for Theoretical
Physics and the organizers of the KITP workshop ``Applications of
Gravitational Lensing'' for hospitality. This work began at this KITP
workshop. We thank an anonymous referee for useful comments on the
manuscript. The research was supported in part by the European Community's
Sixth framework Marie Curie Research Training Network Programme,
contract no. MRTN-CT-2004-505183, by the National Science Foundation
under grant no. PHY05-51164, and by the Department of Energy contract
DE-AC02-76SF00515. This work is based on data obtained as
part of the UKIRT Infrared Deep Sky Survey, UKIDSS (www.ukidss.org).
Some of the data presented herein were obtained at the W.M. Keck
Observatory, which is operated as a scientific partnership among the
California Institute of Technology, the University of California and the
National Aeronautics and Space Administration. The Observatory was made
possible by the generous financial support of the W.M. Keck Foundation.
The authors wish to recognize and acknowledge the very significant
cultural role and reverence that the summit of Mauna Kea has always had
within the indigenous Hawaiian community.  We are most fortunate to have
the opportunity to conduct observations from this mountain.
    Funding for the Sloan Digital Sky Survey (SDSS) and SDSS-II has been
provided by the Alfred P. Sloan Foundation, the Participating
Institutions, the National Science Foundation, the U.S. Department of
Energy, the National Aeronautics and Space Administration, the Japanese
Monbukagakusho, and the Max Planck Society, and the Higher Education
Funding Council for England. The SDSS Web site is http://www.sdss.org/.
The SDSS is managed by the Astrophysical Research Consortium (ARC)
for the Participating Institutions. The Participating Institutions are
the American Museum of Natural History, Astrophysical Institute Potsdam,
University of Basel, University of Cambridge, Case Western Reserve
University, The University of Chicago, Drexel University, Fermilab, the
Institute for Advanced Study, the Japan Participation Group, The Johns
Hopkins University, the Joint Institute for Nuclear Astrophysics, the
Kavli Institute for Particle Astrophysics and Cosmology, the Korean
Scientist Group, the Chinese Academy of Sciences (LAMOST), Los Alamos
National Laboratory, the Max-Planck-Institute for Astronomy (MPIA), the
Max-Planck-Institute for Astrophysics (MPA), New Mexico State
University, Ohio State University, University of Pittsburgh, University
of Portsmouth, Princeton University, the United States Naval
Observatory, and the University of Washington.

\end{document}